\documentclass[aps,prb,twocolumn,superscriptaddress,showpacs]{revtex4}
\usepackage{graphicx}
\usepackage{latexsym}
\usepackage{amsmath}
\usepackage{txfonts}
\usepackage{helvet}
\usepackage{braket}
\usepackage{amscd}
\usepackage{mathptmx}

\begin{document}

\title{An orbital-selective spin liquid in a frustrated heavy fermion spinel LiV$_2$O$_4$}

\author{Yasuhiro Shimizu$^*$}
\affiliation{Department of Physics, Graduate School of Science, Nagoya University, Furo-cho, Chikusa-ku, Nagoya 464-8602, Japan}
\author{Hikaru Takeda}
\affiliation{Department of Physics, Graduate School of Science, Nagoya University, Furo-cho, Chikusa-ku, Nagoya 464-8602, Japan}
\author{Moe Tanaka}
\affiliation{Department of Physics, Graduate School of Science, Nagoya University, Furo-cho, Chikusa-ku, Nagoya 464-8602, Japan}
\author{Masayuki Itoh}
\affiliation{Department of Physics, Graduate School of Science, Nagoya University, Furo-cho, Chikusa-ku, Nagoya 464-8602, Japan}
\author{Seiji Niitaka}
\affiliation{RIKEN Advanced Science Institute, 2-1, Hirosawa, Wako, Saitama 351-0198, Japan}
\author{Hidenori Takagi}
\affiliation{RIKEN Advanced Science Institute, 2-1, Hirosawa, Wako, Saitama 351-0198, Japan}

\date{\today}

\begin{abstract}
	{\bf The pronounced enhancement of the effective mass is the primary phenomenon associated with strongly correlated electrons. 
	In the presence of local moments, the large effective mass is thought to arise from Kondo coupling, the interaction between itinerant and localised electrons. 
	However, in $d$ electron systems, the origin is not clear because of the competing Hund's rule coupling. 
	Here we experimentally address the microscopic origin for the heaviest $d$ fermion in a vanadium spinel LiV$_2$O$_4$ having geometrical frustration. 
	Utilising orbital-selective $^{51}$V NMR, we elucidate the orbital-dependent local moment that exhibits no long-range magnetic order despite persistent antiferromagnetic correlations. 
	A frustrated spin liquid, Hund-coupled to itinerant electrons, has a crucial role in forming heavy fermions with large residual entropy. 
	Our method is important for the microscopic observation of the orbital-selective localisation in a wide range of materials including iron pnictides, cobaltates, manganites, and ruthnates. 
	} 

\end{abstract}


\keywords{}

\maketitle

\section*{Introduction}
	Electrons in metals behave as quasiparticle dressing interactions. 
	The mass of quasiparticles often becomes extremely heavy when metallic phases are close to the quantum critical boundary for the insulating or magnetic phase\cite{Steglich}. 
	The microscopic understanding of heavy quasiparticles (HQs) has been a goal of modern many-body quantum statistics. 
	An established route for HQs is antiferromagnetic Kondo coupling between localised $f$ spins and itinerant electrons in rare-earth metals. 
	In contrast to the $f$-electron case, the presence of localised spins is not apparent for $d$-electron systems. 
	Alternative routes driving $d$ HQ formation have been challenging issues in strongly correlated electron physics. 

	A representative $d$ HQ material is the vanadium spinel LiV$_2$O$_4$ (Ref. \onlinecite {Kondo, Urano}), which has a highly-frustrated pyrochlore lattice for the $B$ site V$^{3.5+}$ ($3d^{1.5}$) ions (Fig. 1a). 
	Anisotropic orbital-dependent intersite interactions give an itinerant $e_g^\prime$ orbital and a more localised $a_{1g}$ orbital through a small trigonal distortion of the VO$_6$ octahedron (Fig. 1b)\cite{Anisimov, Nekrasov}. 
	The HQ was initially explained by off-site Kondo exchange interactions, $J_{\rm K}$, between localised $a_{1g}$ moments and itinerant $e_g^\prime$ electrons (Fig. 1b)\cite{Anisimov}. 
	In the $t_{2g}$ manifold, however, the strong on-site ferromagnetic Hund's exchange interaction, $J_{\rm H}$, can overcome $J_{\rm K}$. 
	Many alternative scenarios such as geometrical frustration via antiferromagnetic interactions $J_{\rm AF}$,\cite{Lacroix2, Burdin, Hopkinson}, electron correlations\cite{Kusunose, Arita}, and spin-orbital fluctuations\cite{Yamashita, Hattori} have been proposed. 

	\begin{figure}
	\includegraphics[scale=0.49]{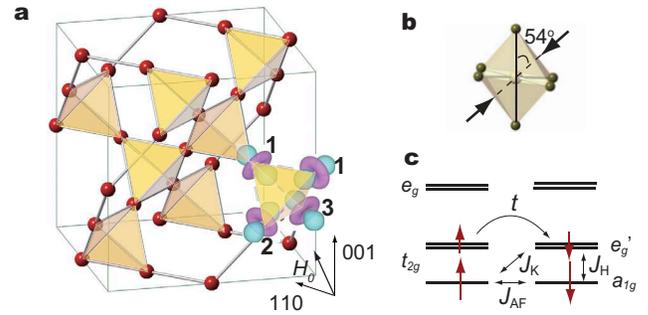} 
	\caption{\label{Fig1} 
	{\bf V pyrochlore lattice and exchange interactions between $3d$ orbitals in LiV$_2$O$_4$.} 
	({\bf a}) Corner-sharing V-tetrahedra offer geometric frustration against magnetic ordering. 
	In the cubic lattice with local trigonal distortion, three vanadium sites with $a_{1g}$-like orbitals, numbered {\bf 1}, {\bf 2}, and {\bf 3} are not equivalent to the magnetic field $H_0$ rotated from the [001] to [110] axis. 
	({\bf b}) A trigonally distorted VO$_6$ ligand field splits a five-fold $3d$ level into $e_g$ and $t_{2g}$ having lower $a_{1g}$ and upper $e_g^\prime$ with 1.5 electrons. 
	For the {\bf 2} and {\bf 3} sites, the trigonal axis is located at $54^\circ$ measured from the crystal [001] axis. 
	({\bf c}) In a Kondo lattice having localised $a_{1g}$ and itinerant $e_g^\prime$ electrons with transfers $t$, exchange interactions between the occupied orbitals are composed of the on-site ferromagnetic Hund's coupling $J_{\rm H}$, the off-diagonal Kondo coupling $J_{\rm K}$, and the off-site antiferromagnetic kinetic exchange $J_{\rm AF}$. 
	}
	\end{figure}

	Experimentally, the interpretations have been unclear for HQs in LiV$_2$O$_4$. 
	Charge-sensitive probes such as resistivity\cite{Kondo}, photoemission\cite{Shimoyamada}, and optical\cite{Takenaka} measurements showed crossover from a high-temperature incoherent metal to a low-temperature Fermi liquid across the characteristic temperature $T^* \sim$ 20-30 K. 
	In contrast, spin-sensitive probes including static spin susceptibility \cite{Kondo, Mahajan} and inelastic neutron scattering\cite{Lee} measurements imply local moments with antiferromagnetic correlations at low temperatures. 
	Furthermore anomalous temperature $T$ dependences on the specific heat $C/T$ and the Hall coefficient conflict with those expected in a conventional Fermi liquid\cite{Kondo}. 
	Despite the theoretical view of orbital-selective interactions, no experimental effort has been made to detect the orbital degrees of freedom. 

	Here, we address the first experimental approach for microscopic observations of the $d$ HQ via orbital-resolved nuclear magnetic resonance (ORNMR) measurements in LiV$_2$O$_4$. 
	The previous NMR experiments using closed-shell Li sites\cite{Kondo, Mahajan} only measured the net spin susceptibility proportional to the bulk value because the hyperfine interactions at the Li site surrounded by 12 vanadium sites average out the anisotropy. 
	Our ORNMR spectroscopic approach using on-site $^{51}$V spins on a high-quality single crystal is sensitive to the orbital-dependent local spin susceptibility, which is beneficial for
probing strongly correlated electrons with the orbital degrees of freedom. 

\section*{Results}
{\bf ORNMR Knight shift.} 
	In LiV$_2$O$_4$ the spin susceptibility $\chi^{\rm s}$ consists of the $a_{\rm 1g}$ and $e_g^{\prime}$ components: $\chi^{\rm s} = \chi^a + \chi^e$ (hereafter the superscripts $a$ and $e$ denote $a_{\rm 1g}$ and $e_g^{\prime}$, respectively). 
	The NMR frequency shift called the Knight shift, ${\bf K}^{\rm s} = (K^{\rm s}_x, K^{\rm s}_y, K^{\rm s}_z)$ ($x$, $y$, and $z$ are the principal axes), measures the spin susceptibility via the hyperfine interaction $\mathcal {H}_{\rm n} =  \sum_i{\bf I} \cdot {\bf A}_i \cdot {\bf \tilde{s}}_i$ with the hyperfine coupling tensor {\bf A}$_i$ between the nuclear spin {\bf I} and the paramagnetic spin polarisation ${\bf \tilde{s}}_i$ for the $i$-th electron under an external magnetic field. 
	Whereas the isotropic shift $K^{\rm s}_{\rm iso}= (K^{\rm s}_z+2K^{\rm s}_x)/3$ due to the core polarization is proportional to $\chi^{\rm s}$, the anisotropic part $K^{\rm s}_{\rm ax} = 2(K^{\rm s}_z-K^{\rm s}_x)/3$ due to the orbital-specific dipole hyperfine interaction\cite{Abragam} is expressed by using the principal $z$ component of the coupling constants, $A^a_z$ and $A^e_z$, 
	\begin{eqnarray}
	K^{\rm s}_{\rm ax} = \frac{1}{N\mu_{\rm B}}(A^a_z\chi^a + A^e_z\chi^e ), 
	\end{eqnarray} 
	where $N$ is Avogadro's number and $\mu_{\rm B}$ is the Bohr magneton. 
	In contrast to $K^{\rm s}_{\rm iso}$, $K^{\rm s}_{\rm ax}$ measures the hyperfine-weighted average of the orbital-dependent spin susceptibility. 
	In the ionic limit, $A^a_z$ and $A^e_z$ are given by a quadratic combination of the angular momentum (see Methods) with the reversed sign and the same amplitude, $A^a_z = -A^e_z > 0$ (Ref. \onlinecite{Abragam}). 
	Hence, we can distinguish which orbital dominates the spin susceptibility from the sign of $K^{\rm s}_{\rm ax}$ and obtain the orbital occupation from the amplitude. 
	Namely, $K^{\rm s}_{\rm ax}$ should be positive (negative) for $\chi^a > \chi^e$ ($\chi^a < \chi^e$). 

	\begin{figure}
	\includegraphics[scale=0.45]{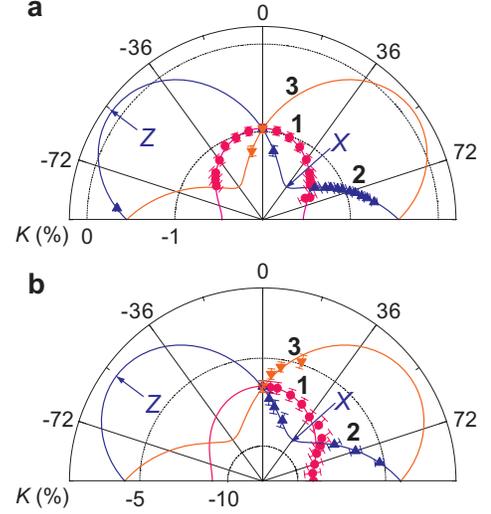}
	\caption{\label{Fig2} 
	{\bf Angle dependence of the $^{51}$V Knight shift $K$ of LiV$_2$O$_4$ at (a) 300 K and (b) 2 K.} 
	$K$ is plotted against the magnetic field direction measured from the [001] axis to the orthogonal [110] axis. 
	The $K$ data with different colors and symbols arise from three vanadium sites {\bf 1}, {\bf 2}, and {\bf 3} due to different orbital configurations against the external magnetic field and trace three curves of the general formula, as shown in Eq. (2). 
	Error bar is defined by a standard deviation of the Lorentzian fit of the NMR spectra. 
	The principal $Z$ and $X$ axes of $K$ for {\bf 2} and {\bf 3} are respectively located at $\mp54^\circ$ and $\pm36^\circ$ measured from [001]. 
	}
	\end{figure}

	The $^{51}$V Knight shift tensors of LiV$_2$O$_4$ are determined from the angle dependence of $K$ for 2-300 K, as shown in Fig. 2. 
	The $^{51}$V NMR spectra were detectable only at specific angles, where the nuclear quadrupole interaction almost vanishes, because the nuclear spin-spin relaxation times at other angles are too fast. 
	The obtained $K$ traces three cosine curves of Eq. (2) in the Methods, which satisfies the cubic $Fd3m$ lattice. 
	The principal $z$ axis of $K$ at $\pm 54^\circ$ for the two V sites indicates the $3d$ orbital symmetry governed by the trigonal VO$_6$ crystal field (Fig. 1b). 
	At 300 and 2 K the relationship $K^{\rm s}_z > K^{\rm s}_x$ ($K^{\rm s}_{\rm ax}>0$) shows the $a_{1g}$-dominant spin susceptibility ($\chi^a > \chi^e$). 
	The result is consistent with the localised nature of the $a_{1g}$ orbital, as theoretically suggested\cite{Anisimov, Lacroix2, Arita, Nekrasov}. 

{\bf Temperature dependence of orbital occupations.} 
	To address HQ formation, we measured the thermal variations of the $^{51}$V Knight shifts $K^{\rm s}_{\rm iso}$ and $-K^{\rm s}_{\rm ax}$ in comparison with the $^{7}$Li Knight shift $^7K^{\rm s}$ and the bulk spin susceptibility $\chi$ (Fig. 3a). 
	Good linearity was observed between these Knight shifts and $\chi$ (Fig. S1). 
	All of data show a Curie-Weiss-like increase at high temperatures, followed by a broad maximum at approximately 20 K. 
	The results agree with the spin susceptibility for high-quality crystals, free from a Curie-tail increase at low temperatures\cite{Kondo, Urano}. 
	The on-site $^{51}$V Knight shift probes the spin susceptibility with greater sensitivity than the off-site $^7$Li one and shows a smooth decrease below $T^*$. 
	Below 5 K $K^{\rm s}$ becomes nearly $T$-independent, as observed in the Fermi liquid. 

	\begin{figure}
	\includegraphics[scale=0.52]{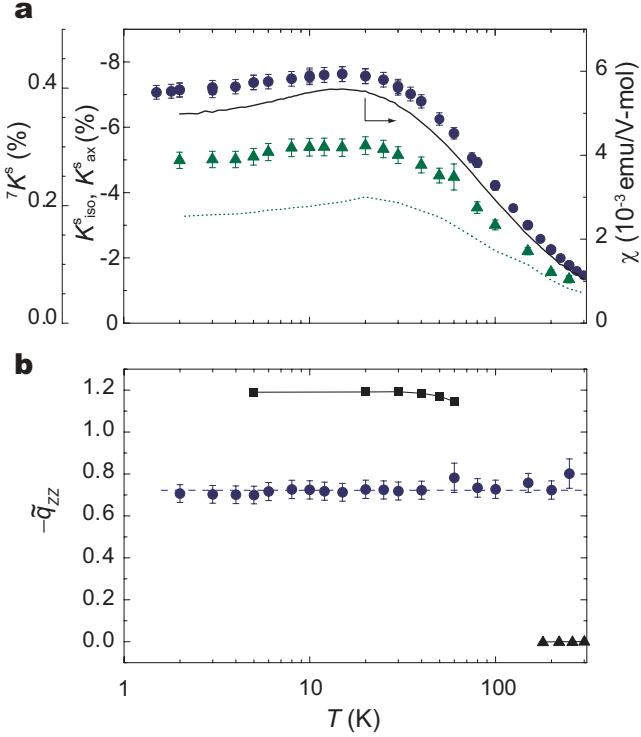}
	\caption{\label{Fig3} 
	{\bf Temperature dependence of spin and orbital polarisations in LiV$_2$O$_4$}. 
	({\bf a}) Local spin susceptibilities probed by $^{51}$V isotropic shift $K^{\rm s}_{\rm iso}$ (circles with error bars), axially anisotropic shift $-K^{\rm s}_{\rm ax}$ (triangles with error bars), and $^{7}$Li Knight shift $^7K^{\rm s}$ (green dotted line), where	error bars are defined by the standard deviation. 
 The right-hand axis shows the bulk susceptibility $\chi$ (solid curve). 
	A constant $K_0 = 0.4\%$ obtained from the $K$-$\chi$ plot (Fig. S1) is subtracted from $K$. 
	({\bf b}) The effective $3d$ orbital polarisation $\tilde{q}_{zz}$ (blue circle with error bars) in comparison with the fully $a_{1g}$ polarised value $\tilde{q}_{zz,0} = -1.2$ in Lu$_2$V$_2$O$_7$ (black square)\cite{Kiyama2006}, and the fully unpolarised value $\tilde{q}_{zz,0} = 0$ in a weakly correlated metal V$_2$O$_3$ (black circle) (Y.S., M.I., $\&$ Y. Ueda, unpublished results). 
	$\tilde{q}_{zz}$ is obtained from $K^{\rm s}_{\rm ax}$ divided by $K^{\rm s}_{\rm iso}$ to cancel the spin polarisation. 
 	}
	\end{figure}

	When $a_{1g}$ local moments becomes Fermi liquid with HQ via $a_{1g}$-$e_g^\prime$ hybridisation or intersite Kondo coupling below $T^*$, $\chi^a$ is expect to decrease significantly, whereas $\chi^a$ is less sensitive. 
	It could lead a decrease of $K^{\rm s}_{\rm ax}$ from Eq. (1). 
	To inspect this property, we plot $K^{\rm s}_{\rm ax}/K^{\rm s}_{\rm iso} \equiv \tilde{q}_{zz}$ against $T$ in Fig. 3b. 
	We find no appreciable change in $\tilde{q}_{zz}$ for 2-300K. 
	This lack of change signifies that the localised character of the $a_{1g}$ orbital persists to the Fermi liquid state across $T^*$. 
	Although $\tilde{q}_{zz}$ of the $f$-electron system has not been reported, the nonlinear relationship in the $K$-$\chi$ plot may be a manifestation of Kondo coupling in the Ce and U-based compounds\cite{Curro, Sakai, Kambe}. 

	$\tilde{q}_{zz}$ reflects the $3d$ orbital polarisation when $\chi^a$ and $\chi^e$ scale to electron occupations. 
	From $\tilde{q}_{zz} = -0.7$ we can evaluate the mixing ratio of the $a_{1g}$ and $e_g^\prime$ orbitals in LiV$_2$O$_4$. 
	The singly occupied $a_{1g}$ orbital has $\tilde{q}_{zz}= -1.2$, as observed in the insulating pyrochlore material Lu$_2$V$_2$O$_7$ (Ref. \onlinecite{Kiyama2006}). 
	In contrast, $\tilde{q}_{zz}$ vanishes for equivalent mixing of $a_{1g}$ and $e_g^\prime$, as observed in a less correlated metal V$_2$O$_3$ (Fig. 3b). 
	The observed intermediate $\tilde{q}_{zz}$ in LiV$_2$O$_4$ manifests a significant $e_g^\prime$ contribution to the spin susceptibility in throughout the temperature range. 
	Namely, the $e_g^\prime$ spin must be polarised via Hund's rule coupling to the localised $a_{1g}$ spins under the magnetic field, although the itinerancy of $e_g^\prime$ is much better than that of $a_{1g}$. 
	The occupation ratio is evaluated as $a_{1g} : e_g^\prime = 4 : 1$ (see Methods), corresponding to the electron numbers of $n_{a_{1g}} \sim 1$ and $n_{e_g^\prime} \sim 0.25$ for $3d^{1.5}$. 
	The half-filling $a_{1g}$ occupation is distinct from that expected in the tight-binding calculation without electron correlations, where $n_{a_{1g}} : n_{e_g^\prime} = 1 : 4 $ (Ref. \onlinecite{Hattori}). 
	However, it is compatible with the strongly localised $a_{1g}$ picture due to the strong renormalization into the Mott insulating state\cite{Nekrasov, Kusunose, Arita} and provides microscopic evidence for orbital-dependent localisation, which is robust across $T^*$. 
	
{\bf Dynamical spin susceptibility of the orbital-selective spin liquid} 
	Another interesting issue is the dynamical part probed by the nuclear spin-lattice relaxation rate $T_1^{-1}$. 
	$(T_1T)^{-1}$ is generally given by $(T_1T)^{-1} \sim \sum_q |A_\perp ({\bf q})|^2{\rm Im} \chi_{\perp}({\bf q}, \omega)/\omega$ (Ref. \onlinecite{Moriya}), where $A_\perp ({\bf q})$ is the wave vector ${\bf q}$ component of the hyperfine coupling constant normal to the quantization axis, and $\chi_{\perp}({\bf q}, \omega)$ is the transverse dynamical spin susceptibility at the NMR frequency $\omega$. 
	In a cubic lattice, $A_\perp ({\bf q})$ and $\chi_{\perp}({\bf q}, \omega)$ are isotropic for the $^{51}$V and $^7$Li sites. 
	$(T_1T)^{-1}$ measured for $^{51}$V and $^7$Li (Fig. 4a) follows the linear relationship ${(^{51}T_1T)}^{-1} = C {(^7T_1T)^{-1}}$ + $C_0$, where the linear coefficient $C = 1.0\times 10^3$ is close to the square ratio of the hyperfine coupling, $\frac{(^{51}\gamma ^{51}A^{\rm  s}_{\rm iso})^2}{(^7\gamma ^7A^{\rm  s}_{\rm iso})^2} = 1.1 \times 10^3$ and $C_0$ = 64 s$^{-1}$K arises from the $T$-invariant orbital component. 
	The scaling relation indicates that the Li sites probe spin fluctuations via the transferred hyperfine interaction and allows us to evaluate unobservable $(^{51}T_1T)^{-1}$ at low temperatures from $(^7T_1T)^{-1}$. 

	\begin{figure}
	\includegraphics[scale=0.5]{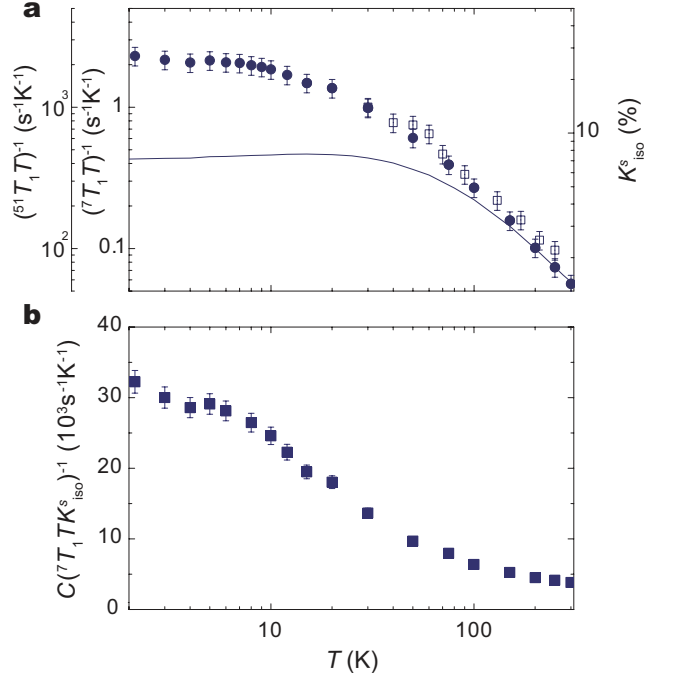}
	\caption{\label{Fig4} 
	{\bf The dynamical spin susceptibility compared to a static one in LiV$_2$O$_4$}. 
	{\bf a}, The temperature dependence of the nuclear spin-lattice relaxation rate divided by temperature $T$, $(T_1T)^{-1}$, obtained for $^7$Li (solid circle with error bars) and $^{51}$V (open square with error bars) NMR (left axes), and the isotropic $^{51}$V Knight shift $K^{\rm s}_{\rm iso}$ (solid line, right axis) in LiV$_2$O$_4$, where error bars are defined by the standard deviation.  
	{\bf b}, $C(^7T_1T)^{-1}$ (solid square with error bars) divided by $K^{\rm s}_{\rm iso}$, where $C(^7T_1T)^{-1}$ is equivalent to the spin part of $(^{51}T_1T)^{-1}$. 
	}
	\end{figure}
	In the present case, $(T_1T)^{-1}$ is governed by paramagnetic fluctuations of local moments at high temperatures. 
	Above 150 K the scaling behaviour between $(^7T_1T)^{-1}$ and $K^{\rm s}_{\rm iso}$ is indeed observed ($C(^7T_1TK^s_{\rm iso})^{-1}$ = constant in Fig. 4b). 
	Below 150 K a progressive $(^7T_1T)^{-1}$ increase indicates antiferromagnetic correlation, consistent with the growth of $\chi({\bf q}, \omega)$ at a finite ${\bf q}$ (= 0.64$\AA^{-1}$) in the inelastic neutron scattering measurements below 80 K (Ref. \onlinecite{Lee}). 
	Therefore the suppression of $\chi^{\rm s}$ at low temperatures likely comes from the short-range antiferromagnetic correlation with the exchange interaction $J_{\rm AF} \sim$ 150 K. 
	Nevertheless, no long-range magnetic ordering occurs down to 1.5 K, the energy scale of $\sim J_{\rm AF}/100$. 
	It suggests that the frustrated $a_{1g}$ spins form in a quantum liquid at low temperatures with low-lying excitations. 

\section*{Discussion}
	Our results provide significant insights into the formation of $3d$ HQ. 
	As observed in the $T$-independent $\tilde{q}_{zz}$, the $a_{1g}$ spins likely remain incoherent, even entering into a coherent 'Fermi liquid' state, and couple ferromagnetically to $e_g^\prime$ spins. 
	No indication of Kondo coupling was observed down to low temperatures despite the large antiferromagnetic fluctuations. 
	The remaining local moments can be highly frustrated and carry large residual entropy\cite{Lacroix2, Burdin, Hopkinson}. 
	The itinerant $e_g^\prime$ electrons interact with the underlying spin liquid via the Hund's rule coupling. 
	Thus the $3d$ HQ behaviour in LiV$_2$O$_4$ could be mapped on the frustrated ferromagnetic Kondo lattice. 

	In the absence of antiferromagnetic Kondo coupling, the HQ formation has not established theoretically. 
	In a Hubbard model calculation, the Kondo-like coherence peak appears on the boundary of the orbital-selective Mott transition for the $a_{1g}$ part\cite{Arita}. 
	Even in such a case, $\chi^a$ may vary across $T^*$, while $\chi^e$ be invariant. 
	Our results suggest that, if the Kondo-like peak appears, a large fraction of the incoherent spins still remains and carries entropy. 
	Such fractionalisation of the nearly localised electron might be common to strongly correlated electron systems\cite{Sachdev} where localised and itinerant characters coexist. 
	Furthermore, the chirality degrees of freedom might provide appreciable entropy in the pyrochlore lattice\cite{Motome} and contribute to the anomalous Hall effect\cite{Urano}. 

	The ORNMR technique offers new holographic experiments that could give microscopic insights into strongly correlated electrons.	
	Various orbital-resolved tools, such as X-ray absorption and photoemission spectroscopy, have been recently developed. 
	The Knight shift measurement has a unique advantage in detecting the orbital-dependent local spin susceptibility via the magnetic hyperfine interactions between $d$ spins and on-site nuclear spins. 
	The method has not been achieved in rare-earth heavy fermion compounds because of the difficulty in detecting NMR signals for the on-site nuclear spins\cite{Curro}. 
	Additional technical improvements in the NMR measurements may reveal the hidden orbital-selective Mott transition in transition metal oxides, such as ruthenates, pnictides, and manganites. 

\section*{Methods}
{\bf NMR measurements.}
	The ORNMR experiments were performed on a single crystal of LiV$_2$O$_4$ synthesised by the self-flux method\cite{Matsushita}. 
	The crystal with the octahedral shape was placed on a two-axis goniometer and rotated under a fixed magnetic field $H_0$ = 9.402 and 8.490 T. 
	The NMR spectra were obtained from spin-echo signals after two $\pi/2$ pulses separated by a time $\tau$. 
	The $^{51}$V NMR measurements were made only for powder samples above 50 K (Ref. \onlinecite{Mahajan}) likely due to the fast spin-echo decay time $T_2$ at low temperatures. 
	To overcome this problem, we used a short $\tau = 3-10$ $\mu$s and a magnetic field precisely ($<$0.1$^\circ$) aligned to the crystal axis equivalent to the magic angle of the nuclear quadrupole interaction. 
	Otherwise, the NMR signals were depressed owing to the fast $T_2$. 

	The angular dependence of the $^{51}$V Knight shift $K(\theta)$ with the local trigonal symmetry is fitted into the general formula\cite{Volkov}
	\begin{eqnarray}
	K = K_{\rm iso} - \tfrac{(K_z - K_x)}{6}{\rm sin}2\theta
	\end{eqnarray} 
	for the V1 site and 
	\begin{eqnarray}
	K = K_{\rm iso} \pm \tfrac{(K_z - K_x)}{2}{\rm cos}(2\theta \pm54.7^\circ)
	\end{eqnarray} 
	for the V2 or V3 site related by a mirror symmetry, where $K_{\rm iso} = \tfrac{K_z+2K_x}{3}$. 

{\bf Magnetic hyperfine interactions.}
	Magnetic hyperfine interactions between $3d$ electron spins and nuclear spins, and the analysis of the $^{51}$V Knight shift in LiV$_2$O$_4$. 
	Magnetic hyperfine interactions in $3d$ systems are generally given by\cite{Abragam}
\onecolumngrid
	\begin{equation}	
	\mathcal {H}_{\rm n} = \mathcal {P}\sum_i \left[ {\bf l}_i \cdot {\bf I} - \kappa {\bf s}_i \cdot {\bf I} - \tfrac{2}{(2l_i-1)(2l_i + 3)}\left \{ \tfrac{3}{2}({\bf l}_i\cdot {\bf s}_i)({\bf l}_i\cdot {\bf I}) + \tfrac{3}{2}({\bf l}_i\cdot {\bf I})({\bf l}_i\cdot {\bf s}_i) - l_\alpha(l_i + 1)({\bf I}\cdot {\bf s}_i) \right \} \right]  
	\end{equation} 
\twocolumngrid
	where ${\bf l}_i$, ${\bf s}_i$, and {\bf I} denote operators of orbital and spin of the $i$-th electron, and the nuclear spin, respectively, the coefficient $\mathcal {P} = 2\mu_{\rm B}\gamma_{\rm n}\hbar \braket {r^{-3}}$ using the Bohr magneton $\mu_{\rm B}$, the nuclear gyromagnetic ratio $\gamma_{\rm n}$, the Plank's constant $\hbar$, and a radial expectation value $\braket {r^{-3}}$. 
	The first term represents the orbital contribution that quenches in crystals but partly revives via the Van-Vleck process under the magnetic field. 
	The second term arises from a Fermi contact interaction due to the core polarisation of inner $s$ spins, giving the isotropic hyperfine coupling constant $A^{\rm s}_{\rm iso} = -\mathcal {P} \kappa \delta_{ij}$, where $\kappa$ is a dimensionless parameter ($\kappa \sim 0.5$ for vanadates$^{18}$). 
	The third term denotes anisotropic dipole interactions determined by $3d$ orbital occupations, where the principal components are expressed as $A^{\rm dip}_{\alpha \beta} = 2\mathcal {P} (-q_{\alpha \beta} + \lambda \Lambda^\prime_{\alpha \beta})/7$ by using the equivalent operator of $3d$ angular momentum, 
	\begin{equation}
	q_{\alpha \beta} = \frac{1}{2}(l_\alpha l_\beta+l_\beta l_\alpha)-\frac{1}{3}l(l+1)\delta_{\alpha \beta},  
	\end{equation} 
	with the spin-orbit coupling parameter $\lambda$, and the second-order matrix elements between the ground and excited states, $\Lambda^\prime_{\alpha \beta}$. 
	In the $LS$-coupling the sum of the terms for several electrons can be replaced by 
\onecolumngrid
	\begin{equation}
	\mathcal {H}_{\rm n} = \mathcal {P} {\bf L} \cdot {\bf I} - \mathcal {P}\kappa {\bf S} \cdot {\bf I} - \mathcal {P} \left[\frac{3}{2}\xi \left \{ ({\bf L}\cdot {\bf S})({\bf L}\cdot {\bf I}) + ({\bf L}\cdot {\bf I})({\bf L}\cdot {\bf S})\right \} - \xi L(L+1)({\bf I}\cdot {\bf S}) \right], 
	\end{equation} 
\twocolumngrid
	where {\bf L} and {\bf S} are total orbital and spin, respectively, and $\xi = \frac{2l+1-4S}{S(2l-1)(2l+3)(2L-1)} = \frac{2}{21}$ for $3d^1$. 

	The experimental observable is the Knight shift tensor ${\bf K} = (K_x, K_y, K_z)$ defined as the resonance frequency shift due to the hyperfine interaction of Eq. (4). 
	The spin component $K^{\rm s}$ is obtained by subtracting the small orbital component $K_0$ including the chemical shift and the Van-Vleck shift from the $K$-$\chi$ plot in Fig. S1. 
	For a paramagnetic system, ${\bf s}_i$ in Eq. (4) is replaced by the effective electron spin polarization ${\bf \tilde{s}}_i$ proportional to spin susceptibility $\chi^{\rm s}$. 
	In a multi-orbital system, $\chi^{\rm s}$ is composed of orbital-dependent spin susceptibilities. 
	Then $K^{\rm s}$ is expressed by the sum of hyperfine fields from $3d$ spins. 
	Whereas the isotropic part of $K^{\rm s}$, $K^{\rm s}_{\rm iso} \equiv (2K^{\rm s}_x + K^{\rm s}_z)/3$, is given by 
	\begin{equation}
	K^{\rm s}_{\rm iso} = -\frac{\kappa}{N\mu_{\rm B}}\mathcal {P}\chi^{\rm s}, 
	\end{equation} 
	the anisotropic part, $K^{\rm s}_{\rm ax} \equiv 2(K^{\rm s}_z - K^{\rm s}_x)/3$, is expressed as the arithmetic average of the orbital-dependent spin susceptibility weighted by the principal $z$ component of the hyperfine coupling tensor, as shown in Eq. (1). 
	In LiV$_2$O$_4$ with a local trigonal distortion, 1.5 electrons are filled in two orbitals, $a_{1g}$ and $e_g^\prime$, whose $q_{\alpha \beta}$ are equivalent to those of $d_{3z^2-r^2}$ and $d_{x^2-y^2}$, respectively, taking the principal axes along the trigonal axis: $q_{xx} = q_{yy} = 1$, $q_{zz} = -2$ for $(a_{1g})^1$, while the values are numerically the same but reversed in sign for $(e_g^{\prime})^2$. 
	$q_{\alpha \alpha}$ vanishes when the two orbitals are equally occupied. 
	Using the relation $A_z^a = -A_z^e (= -\tfrac{4}{7}\mathcal {P})$, $K^{\rm s}_{\rm ax}$ is expressed as
	\begin{equation}
	K^{\rm s}_{\rm ax} \simeq  \frac{1}{N\mu_{\rm B}}A^a_z(\chi^a - \chi^e ). 
	\end{equation} 
	The good linearity in $K$-$\chi$ plots indicates $\chi^{\rm s} = \chi^a + \chi^e \sim f\chi^{\rm s} + (1-f)\chi^{\rm s}$, where $f$ is the fraction of $\chi^a$ in $\chi^{\rm s} $. 
	Then $K^{\rm s}_{\rm ax}$ can be further reduced to
	\begin{equation}
	K^{\rm s}_{\rm ax} = -\frac{4}{7N\mu_{\rm B}} \mathcal {P}\chi^{\rm s}(2f-1). 
	\end{equation}
	for the negligible $\lambda \Lambda^\prime_{ij}$ as expected from the small $K_0$. 
	To experimentally obtain the effective orbital polarization $\tilde{q}_{zz}$, we take a ratio of Eqs. (7) and (9) and cancel out the numerical constants and $\chi^{\rm s}$. 
	Namely, 
	\begin{equation}
	\frac{K^{\rm s}_{\rm ax}}{K^{\rm s}_{\rm iso}} \simeq -\frac{8}{7}(2f-1) \equiv \tilde{q}_{zz}. 
	\end{equation} 
	Here $\tilde{q}_{zz} = -8/7$ for a fully $a_{1g}$ polarized case ($f$ = 1), close to the experimentally obtained $\tilde{q}_{zz} \simeq  -1.2$ in LuV$_2$O$_7$ (Ref. 18). 
	From the experimental result $\tilde{q}_{zz} \sim -0.7$ in LiV$_2$O$_4$, we obtained $f \sim$ 0.8, corresponding to the occupation ratio $a_{1g} : e_g^\prime = 4 : 1$ and hence $n \sim  1$ for $a_{1g}$ and $n \sim  0.25$ for $e_g^\prime$. 

{\bf Electric hyperfine interactions} 
	The electrostatic hyperfine interaction can be a direct probe for $3d$ orbital order. 
	In the presence of the anisotropic electric field gradient around the nuclear spin,  the $^{51}$V NMR spectrum is split into seven lines for $I = 7/2$. 
	The NMR spectrum becomes sharpest at [001], identifying the magic angle where the nuclear quadrupole splitting frequency $\delta \nu$ vanishes. 
	Then $\delta \nu$ should have a maximum at $\theta_0 = 54.7^\circ$ satisfying $\delta \nu \sim(3$cos$^2\theta_0 - 1$) = 0, which exactly equivalent to the local trigonal symmetry. 
	We observed a quadrupole splitting frequency $\delta \nu_x$ = 90 kHz in the $^{51}$ V NMR spectra at $H_0 \parallel [110]$, by using a very short pulse interval time $\tau$ = 3 $\mu$s. 
	From the lattice symmetry, we can obtain $\nu_Q \equiv 2\delta \nu_x$ = 180 kHz. 
	We confirmed that $\delta \nu_x$ was independent of temperature down to 2 K and hence the orbital occupation was invariant across $T^*$. 

\section*{Acknowledgements}
	We thank technical assistance by S. Inoue, and valuable discussion with N. Kawakami, Y. Motome, S. Watanabe, T. Tsunetsugu, and S. Sachdev. 
	This work was financially supported by the Grant-in-Aid for Scientific Research on the Priority Area, Novel State of Matter Induced by Frustration, (No. 22014006) from the MEXT, and the Grants-in-Aid for Scientific Research (No. 22684018 and 24340080) from the JSPS. 
	%



\end{document}